# Understanding the Tracking Errors of Commodity Leveraged ETFs

Kevin Guo and Tim Leung

**Abstract** Commodity exchange-traded funds (ETFs) are a significant part of the rapidly growing ETF market. They have become popular in recent years as they provide investors access to a great variety of commodities, ranging from precious metals to building materials, and from oil and gas to agricultural products. In this article, we analyze the tracking performance of commodity leveraged ETFs and discuss the associated trading strategies. It is known that leveraged ETF returns typically deviate from their tracking target over longer holding horizons due to the so-called volatility decay. This motivates us to construct a benchmark process that accounts for the volatility decay, and use it to examine the tracking performance of commodity leveraged ETFs. From empirical data, we find that many commodity leveraged ETFs underperform significantly against the benchmark, and we quantify such a discrepancy via the novel idea of *realized effective fee*. Finally, we consider a number of trading strategies and examine their performance by backtesting with historical price data.

## 1 Introduction

The advent of commodity exchange-traded funds (ETFs) has provided both institutional and retail investors with new ways to gain exposure to a wide array of commodities, including precious metals, agricultural products, and oil and gas. All commodity ETFs are traded on exchanges like stocks, and many have very high liquidity. For example, the SPDR Gold Trust ETF (GLD), which tracks the daily London gold spot price, is the most traded commodity ETF with an average trading volume of 8 million shares and market capitalization of US $31 billion in 2013.[1]

Within the commodity ETF market, some funds are designed to track a constant multiple of the daily returns of a reference index or asset. These are called leveraged ETFs (LETFs). An LETF maintains a constant leverage ratio by holding a variable portfolio of

Kevin Guo
Industrial Engineering & Operations Research (IEOR) Department, Columbia University, New York, NY 10027, e-mail: klg2138@columbia.edu.

Tim Leung
Industrial Engineering & Operations Research (IEOR) Department, Columbia University, New York, NY 10027, e-mail: tl2497@columbia.edu. Corresponding author.

[1] According to ETF Database website (http://www.etfdb.com/compare/volume).

assets and/or derivatives, such as futures and swaps, based on the reference index. For example, the Dow Jones U.S. Oil & Gas Index (DJUSEN) or the Dow Jones U.S. Basic Materials Index (DJUSBM) and their associated ETFs track the stocks of a basket of commodities producers, as opposed to the physical commodity prices. On the other hand, most LETFs are based on total return swaps and commodity futures. The most common leverage ratios are $\pm 2$ and $\pm 3$, and LETFs typically charge an expense fee. Major issuers include ProShares, iShares, VelocityShares and PowerShares (see Table 1). For example, the ProShares Ultra Long Gold (UGL) seeks to return 2x the daily return of the London gold spot price minus a small expense fee. One can also take a bearish position by buying shares of an LETF with a negative leverage ratio. The ProShares Ultra Short Gold (GLL) is an inverse LETF that tracks -2x the daily return of the London gold fixing price. LETFs are a highly accessible and liquid instrument, thereby making them attractive instruments for traders who wish to gain leveraged exposure to a commodity without borrowing money or using derivatives.

For a long LETF, with a leverage ratio $\beta > 0$, the fund must add to a winning position in a bull market to maintain a constant leverage ratio. On the other hand, during a bear market, the fund must sell its losing positions to maintain the same leverage ratio. Similar arguments can be made for short (or inverse) LETFs ($\beta < 0$). As a consequence, LETFs can potentially outperform $\beta$ times its reference during periods of market trending. However, should the LETF exhibit high volatility but no significant movement in price over a period of time, the constant daily re-balancing would cause the fund to decline in value. Therefore, LETFs can be viewed as long momentum but short volatility, and the value erosion due to realized variance of the reference is called *volatility decay* (see [2, 3, 4]). This raises the important question of how well do LETFs perform over a long horizon.

Since their introduction to the market, LETFs a number of criticisms from both practitioners and regulators.[2] Some are concerned that the returns of LETFs exhibit some discrepancies from the goals stated in their prospectuses. In fact, some issuers provide warnings that LETFs are unsuitable for long-term buy-and-hold investors.

Many existing studies focus on equity-based ETFs and their leveraged counterparts. For example, Avellaneda and Zhang [2] study the price behavior and discuss the volatility decay of equity LETFs in different sectors. They find minimal 1-day tracking errors among the most liquid equity ETFs. They explain that an equity LETF can replicate the leveraged returns of its reference through a dynamic portfolio consisting of the component equities.

In contrast, commodities are unique because the physical assets cannot be stored easily. As such, ETF issuers are required to replicate through either warehousing[3], which is very costly, and thus uncommon except for precious metals such as silver and gold, or trading futures with multiple counterparties (see [5]). Since the reference indices may represent the spot prices of physical commodities, futures-based commodity ETFs may fail to track their reference indices perfectly and their tracking performance is subject to the fluctuation and term structure of futures prices. On top of that, most commodity LETFs use over-the-counter (OTC) total return swaps with multiple counterparties to generate the required leverage ratios. The lower liquidity of OTC contracts and counterparty risk can contribute to additional tracking errors. As we show in this paper, tracking errors can seriously affect the long-term fund performance of LETFs.

---

[2] In 2009, the SEC and FINRA issued an alert on the risk of leveraged ETFs on `http://www.sec.gov/investor/pubs/leveragedetfs-alert.htm`.

[3] For more details on the issue of storage cost for commodity ETFs, we refer to the Morningstar Report: "An Ugly Side to Some Commodity ETFs" by Bradley Kay, August 19, 2009.

In a related work, Murphy and Wright [12] perform a $t$-test based on 1-day returns to determine if any commodity LETF has a non-zero tracking error. They conclude that all LETFs have a very good daily tracking performance. However, they do not conduct the analysis over a longer horizon, or account for the volatility decay. There is also no discussion of trading strategies there. On the other hand, Guedj et al. [5] discuss the difficulties faced by an ETF provider in replicating a commodity index using futures. In particular, they point out that the term structure of futures may lead to large deviations between the ETF price and the spot price of a commodity.

In this paper, we analyze the tracking performance of commodity leveraged ETFs. Through a series of regression analyses, we illustrate how the returns of commodity LETFs deviate from the reference returns multiplied by the leverage ratio over different holding periods. In particular, the average tracking error tends to turn more negative over a longer horizon and for higher leveraged ETFs. With in mind that realized variance of the reference can erode the LETF value, we examine the over/under-performance of LETFs with respect to a benchmark that incorporates the effect of volatility decay. From empirical data, we find that many commodity leveraged ETFs in our study underperform significantly against the benchmark, and we quantify such a discrepancy by introducing the *realized effective fee*. Finally, we consider a static trading strategy that involves shorting two LETFs with leverage ratios of different signs, and study its performance and dependence on the realized variance of the reference. We find that the resulting portfolio is always long realized variance both theoretically and empirically, but is also exposed to the tracking errors associated with the two LETFs. We also backtest the strategy through examining its empirical returns over rolling periods.

The rest of the paper is organized as follows. In Section 2, we analyze the returns of commodity LETFs over different holding periods and illustrate horizon dependence of tracking errors. In Section 3, we use a benchmark process that incorporates the realized variance of the reference to study the over/under-performance of each LETF. In Section 4, we discuss a static trading strategy and backtest using historical data. Section 5 concludes the paper and points out a number of directions for future research.

## 2 Analysis of Tracking Error

We first compare the returns of LETFs and their reference indices. For every ETF, we obtain its closing prices and reference index values from Bloomberg for the period December 2008-May 2013. We then calculate the $n$-day returns from $n = \{1, 2, \ldots, 30\}$ using disjoint successive periods (e.g. the return over days 1-30 then returns over days 31-60 for 30-day returns). Let $L_t$ be the price of an LETF and $S_t$ be the reference index value at time $t$. For a given leverage ratio $\beta$, we compare the log-returns of the LETF to $\beta$ times the log-returns of the corresponding reference index. This leads us to define the $n$-day tracking error at time $t$ by

$$Y_t^{(n)} = \ln \frac{L_{t+n\Delta t}}{L_t} - \beta \ln \frac{S_{t+n\Delta t}}{S_t}, \tag{1}$$

where $\Delta t$ represents one trading day. We explore the empirical distribution of the $n$-day tracking error, and then analyze the effect of holding horizon on the magnitude of tracking errors. We remark there are alternative ways to define tracking errors for ETFs. For example, one can consider the difference in relative returns as opposed to log-returns, or the root mean square of the daily differences (see [10]).

| LETF | Reference | Underlying | Issuer | $\beta$ | Fee | Inception |
|---|---|---|---|---|---|---|
| SLV | SLVRLN | Silver Bullion | iShares | 1 | 0.50% | 04/21/2006 |
| AGQ | SLVRLN | Silver Bullion | ProShares | 2 | 0.95% | 12/01/2008 |
| ZSL | SLVRLN | Silver Bullion | ProShares | -2 | 0.95% | 12/01/2008 |
| USLV | SPGSSIG | Silver Bullion | VelocityShares | 3 | 1.65% | 10/13/2011 |
| DSLV | SPGSSIG | Silver Bullion | VelocityShares | -3 | 1.65% | 10/14/2011 |
| GLD | GOLDLNPM | Gold Bullion | iShares | 1 | 0.40% | 11/18/2004 |
| UGL | GOLDLNPM | Gold Bullion | ProShares | 2 | 0.95% | 12/01/2008 |
| GLL | GOLDLNPM | Gold Bullion | ProShares | -2 | 0.95% | 12/01/2008 |
| UGLD | SPGSGCP | Gold Bullion | VelocityShares | 3 | 1.35% | 10/13/2011 |
| DGLD | SPGSGCP | Gold Bullion | VelocityShares | -3 | 1.35% | 10/14/2011 |
| IYE | DJUSEN | Oil & Gas | iShares | 1 | 0.48% | 06/12/2000 |
| DDG | DJUSEN | Oil & Gas | ProShares | -1 | 0.95% | 06/10/2008 |
| DIG | DJUSEN | Oil & Gas | ProShares | 2 | 0.95% | 01/30/2007 |
| DUG | DJUSEN | Oil & Gas | ProShares | -2 | 0.95% | 01/30/2007 |
| DBO | DBOLIX | WTI Crude Oil | PowerShares | 1 | 0.75% | 01/05/2007 |
| UCO | DJUBSCL | WTI Crude Oil | ProShares | 2 | 0.95% | 11/24/2008 |
| SCO | DJUBSCL | WTI Crude Oil | ProShares | -2 | 0.95% | 11/24/2008 |
| UWTI | SPGSCLP | WTI Crude Oil | VelocityShares | 3 | 1.35% | 02/06/2012 |
| DWTI | SPGSCLP | WTI Crude Oil | VelocityShares | -3 | 1.35% | 02/06/2012 |
| IYM | DJUSBM | Building Materials | iShares | 1 | 0.48% | 06/12/2000 |
| SBM | DJUSBM | Building Materials | ProShares | -1 | 0.95% | 03/16/2010 |
| UYM | DJUSBM | Building Materials | ProShares | 2 | 0.95% | 01/30/2007 |
| SMN | DJUSBM | Building Materials | ProShares | -2 | 0.95% | 01/30/2007 |

Table 1: A summary of the 23 LETFs studied in this paper, arranged by commodity type and then leverage. Notice that the non-leveraged (1x) ETFs have the smallest expense fees, and LETFs with higher absolute leverage ratios, $|\beta| \in \{2,3\}$, tend to have higher expense fees. Finally, notice that higher $\beta$ LETFs are much more recent additions to the market.

## *2.1 Regression of Empirical Returns*

We conduct a regression between log-returns of the LETF and its reference index based on the linear model:

$$\ln \frac{L_t}{L_0} = \hat{\beta} \ln \frac{S_t}{S_0} + \hat{c} + \varepsilon, \tag{2}$$

where $\varepsilon \sim N(0, \sigma^2)$ is independent of the reference index value $S_t$, $\forall t \geq 0$. In other words, we run an ordinary least square 1-variable regression between the log-returns for every fixed horizon of $n$ days. Then, we increase the holding period from 1 to 30 days, and observe how the regression coefficients vary.

We display the regression results in Figures 1 through 4 for log-returns over periods of 1, 5, 10, and 20 days. To avoid dependence among returns, we use disjoint time intervals to calculate returns. For example, we use $\frac{S_{20}}{S_0}, \frac{S_{40}}{S_{20}} \dots$ and $\frac{L_{20}}{L_0}, \frac{L_{40}}{L_{20}} \dots$ for 20-day log-returns as the inputs for the regression.

In Figure 1, the regression coefficient $\hat{\beta}$ for DIG ($\beta = 2$, oil & gas) increases from 2 to 2.1 as the holding period lengthens from 1 to 20 days. Although the coefficient of determination $R^2$ is close to 99% for up to 20 days, it is highest for 1-day returns. In Figure 2 for DUG ($\beta = -2$, oil & gas), one again observes $\hat{\beta}$ increasing, and $R^2$ decreasing. For DUG ($\beta = -2$, oil & gas), as $n$ varies from 1 to 20, $\hat{\beta}$ increases from $-2$ to $-1.66$. As a result, this implies that DIG ($\beta = 2$, oil & gas) effectively gains leverage as the holding time increases, while DUG ($\beta = -2$, oil & gas) loses leverage compared to the advertised fund $\beta$.

On the other hand, UGL ($\beta = 2$, gold) and GLL ($\beta = -2$, gold) exhibit very different return behaviors. In Figure 3 the $R^2$ for UGL ($\beta = 2$, gold) is surprisingly worst for the shortest holding period of 1 day, whereas it increases to 95% over a holding period of 20 days. In Figure 4 for GLL ($\beta = -2$, gold), the $R^2$ increases from 35% to 96% when holding the fund from 1 to 20 days. Furthermore, the estimators $\hat{\beta}$ for UGL ($\beta = 2$, gold) and GLL ($\beta = -2$, gold) both slowly approach their advertised $\beta = \pm 2$. The variation of $\hat{\beta}$ for DIG ($\beta = 2$, oil & gas) and UGL ($\beta = 2$, gold) over different holding periods is summarized in Figure 5.

We observe that LETFs that track an illiquid reference, such as the gold bullion index GOLDLNPM, tend to have more tracking errors than those tracking a liquid index, such as the oil & gas index DJUSEN. The oil & gas commodity LETFs involve exchange-traded futures which are liquid proxy to the spot price. The gold and silver bullion LETFs consist of OTC total return swaps. The difficulty and higher costs replication using swaps, as well as infrequent (typically daily) update of the swaps' mark-to-market values can weaken the fund's tracking ability. For example, the 1-day regressions of UGL and GLL ($\beta = \pm 2$, gold) yield $R^2$ values less than 40%, while DIG and DUG ($\beta = \pm 2$, oil & gas ) have 1-day $R^2$ values of over 90%. On the other hand, full physical replication yields the greatest $R^2$, with examples of the non-leveraged gold and silver ETFs, GLD and SLV, respectively. Hence, the replication strategy can significantly affect a fund's tracking errors. A more precise understanding of the effectiveness of swaps, futures, and other replication strategies requires the full holdings history from the ETF provider, which is not publicly available at all times.[4]

In addition, the LETFs we studied have an increasingly negative constant coefficient $\hat{c}$ as the holding time increases. For example, over a holding period of 20-days, DUG ($\beta = -2$, oil & gas) has a 3% decay on returns compared to $\beta$ times its reference index. We would expect this phenomenon, however, since the LETF would need to buy high and sell low, while the reference investor would simply hold his securities. Therefore, the longer the LETF is held, the more likely the fund will underperform against $\beta$ times the reference index. As we will see in Section 3, the constant coefficient $\hat{c}$ depends on two factors, the expense fee charged by the issuer as well as the realized variance of the reference index.

Hence, with this simple linear model for LETF prices, we have observed that although LETFs safely replicate $\beta$ times the reference over short holding periods, they begin to exhibit negative tracking error and deviations in their leverage ratios $\beta$ as the holding time increases. Furthermore, we see that LETFs which attempt to track illiquid spot prices perform much more poorly than expected. We conclude that more factors must be considered when modeling LETF returns.

[4] For a detailed snapshot of the holdings for a proshares ETF, please see `http://www.proshares.com/funds/{XYZ}_daily_holdings.html` where $\{XYZ\}$ is the ETF ticker.

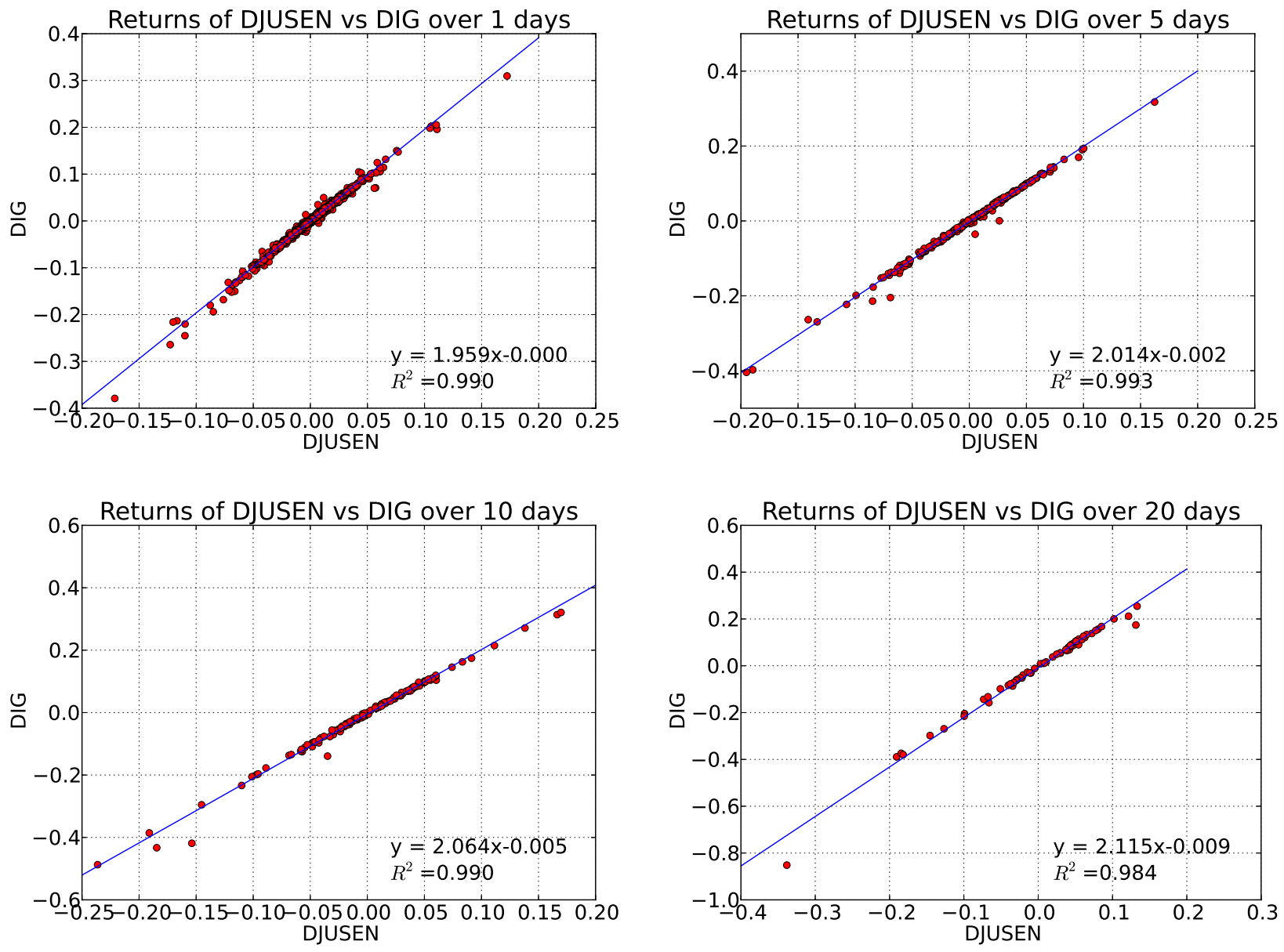

Fig. 1: From top left to bottom right: regression of DJUSEN-DIG ($\beta = 2$, oil & gas) 1, 5, 10, 20-day log-returns. We consider disjoint periods from December 2008 to May 2013.

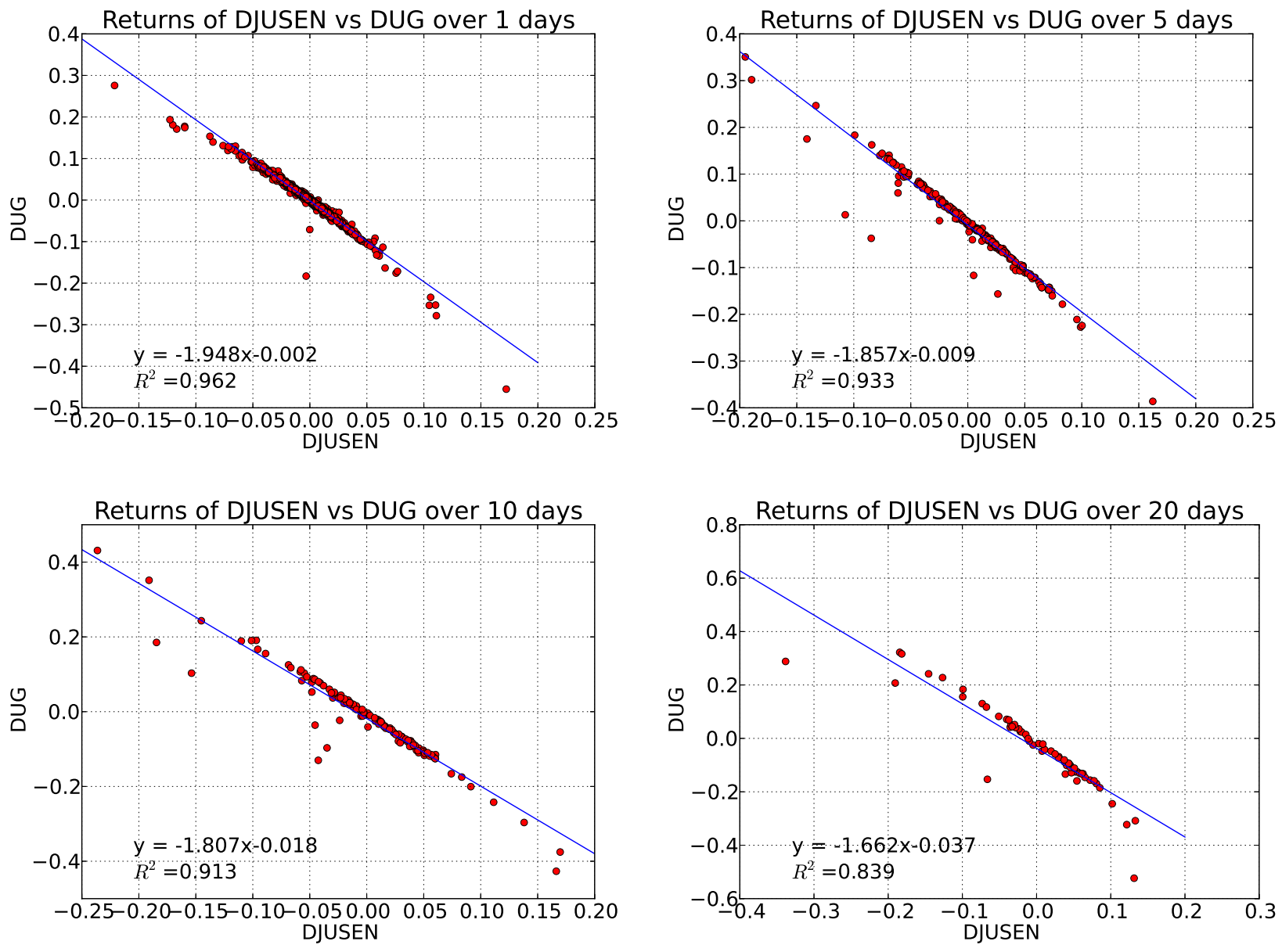

Fig. 2: From top left to bottom right: regression of DJUSEN-DUG ($\beta = -2$, oil & gas) 1, 5, 10, 20-day log-returns. We consider disjoint periods from December 2008 to May 2013.

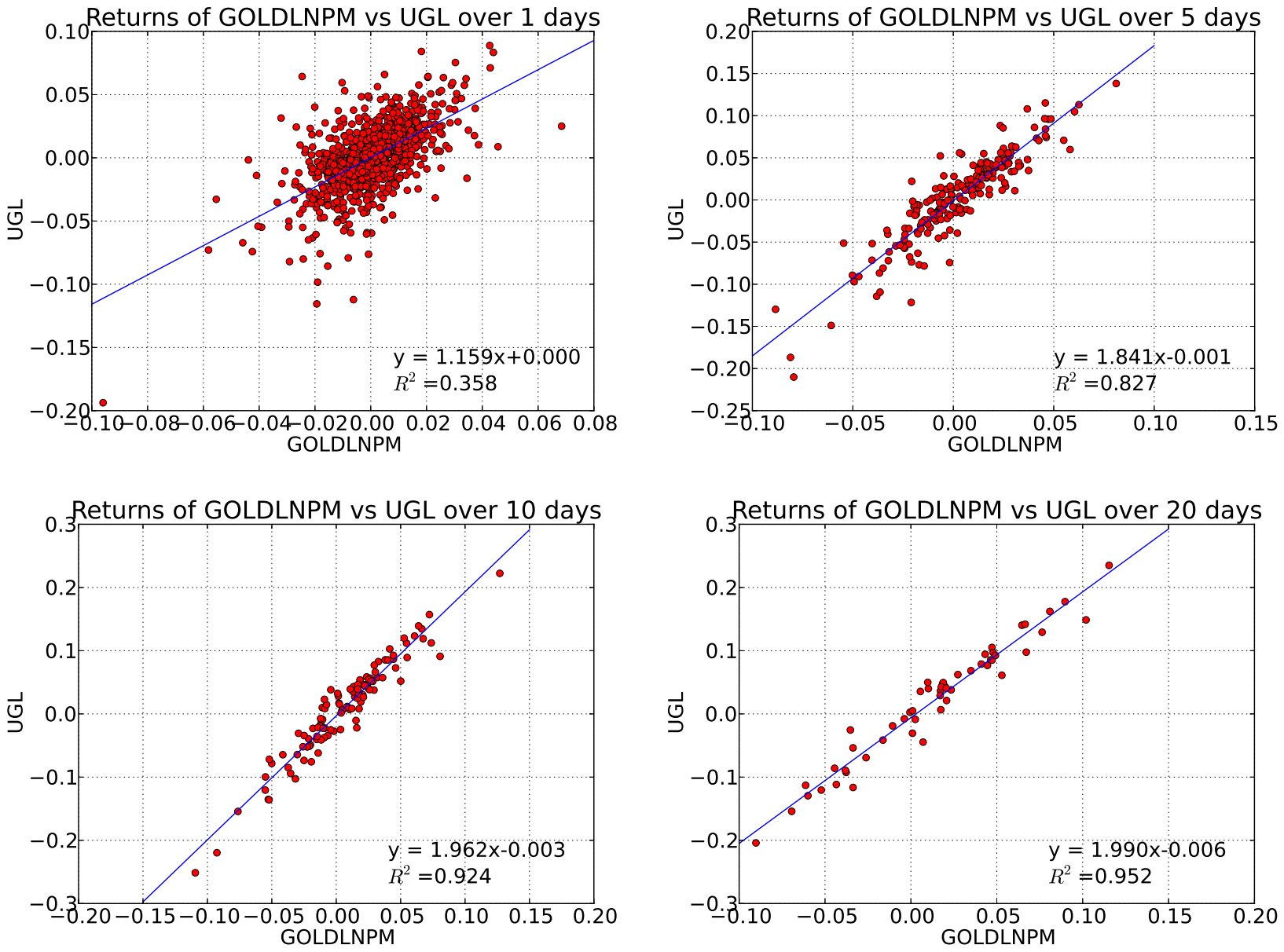

Fig. 3: From top left to bottom right: regression of GOLDLNPM-UGL ($\beta = 2$, gold) 1, 5, 10, 20-day log-returns. We consider disjoint periods from December 2008 to May 2013.

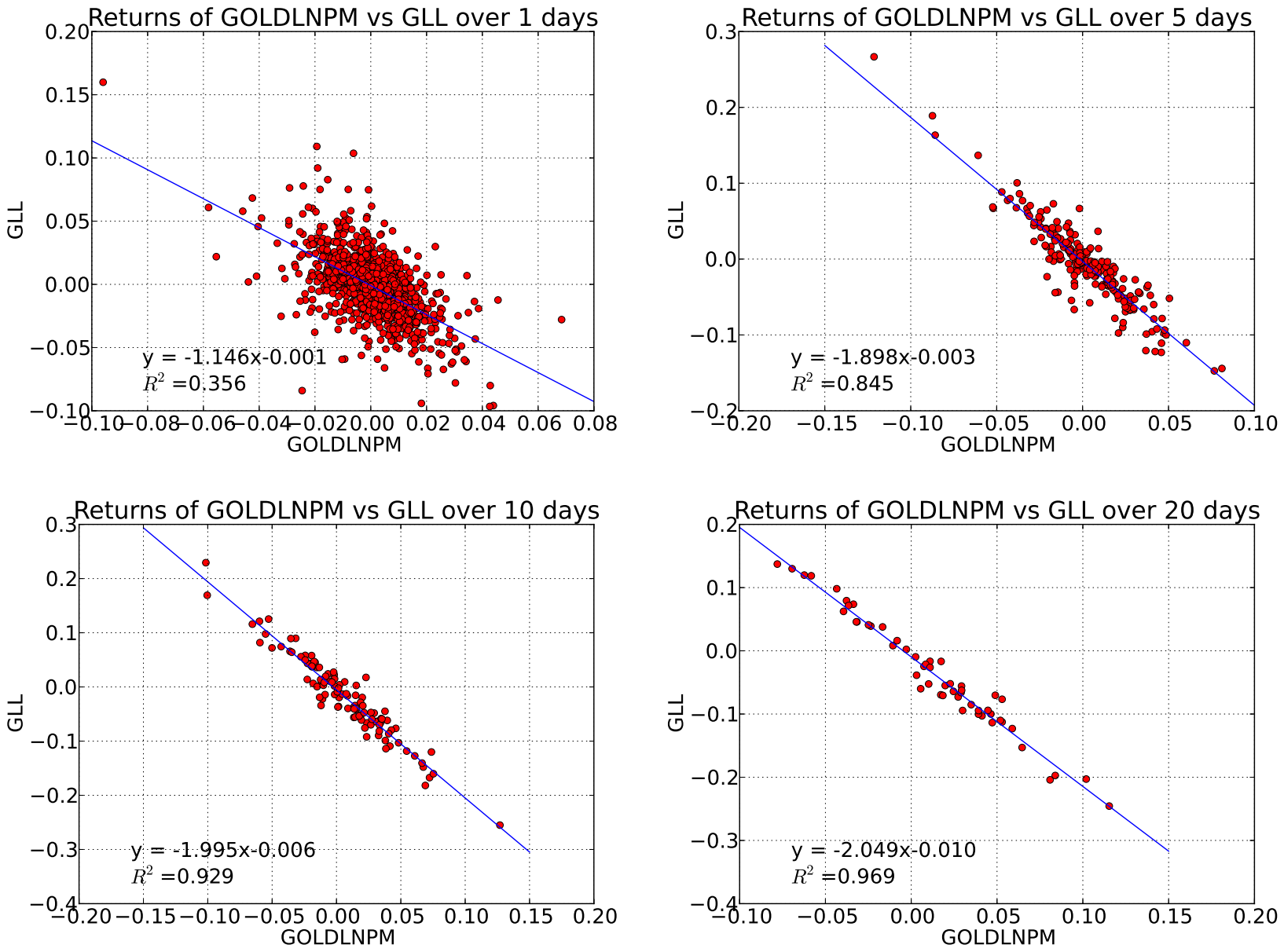

Fig. 4: From top left to bottom right: regression of GOLDLNPM-GLL ($\beta = -2$, gold) 1, 5, 10, 20-day log-returns. We consider disjoint periods from December 2008 to May 2013.

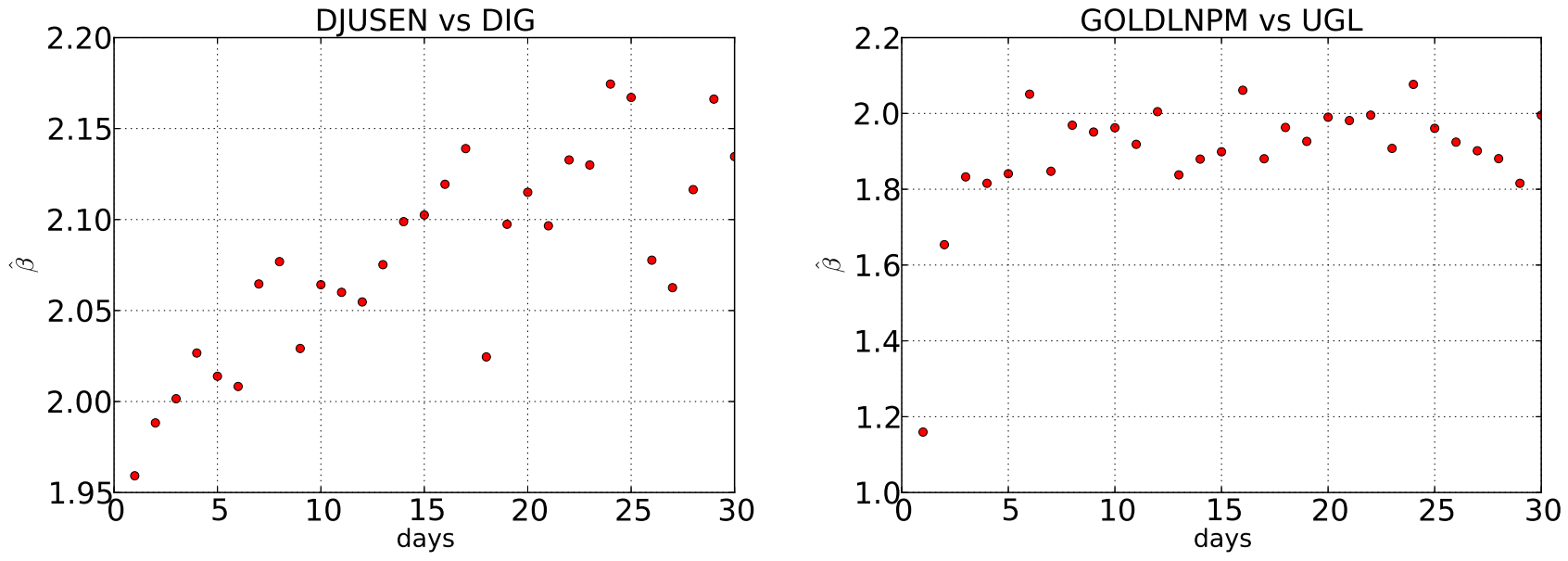

Fig. 5: The estimated $\hat{\beta}$ from the regressions for DJUSEN-DIG ($\beta = 2$, oil & gas), and GOLDLNPM-UGL ($\beta = 2$, gold).

## 2.2 Distribution of Tracking Errors

As defined in (1), the tracking error is the difference between the LETF's log-return and the corresponding multiple of its reference index's log-return. In this section, we examine the distribution of the tracking error. This provides a picture of the LETF's efficiency in its stated goal of replicating the leveraged return of a reference index.

| **LETF** | **Underlying** | $\beta$ | $\mu$ | $\sigma$ |
|---|---|---|---|---|
| SLV | Silver Bullion | 1 | 0.0000 | 0.0302 |
| AGQ | Silver Bullion | 2 | -0.0009 | 0.0539 |
| ZSL | Silver Bullion | -2 | -0.0022 | 0.0543 |
| USLV | Silver Bullion | 3 | -0.0014 | 0.0231 |
| DSLV | Silver Bullion | -3 | -0.0027 | 0.0237 |
| GLD | Gold Bullion | 1 | 0.0000 | 0.0128 |
| UGL | Gold Bullion | 2 | -0.0003 | 0.0221 |
| GLL | Gold Bullion | -2 | -0.0005 | 0.0221 |
| UGLD | Gold Bullion | 3 | -0.0006 | 0.0134 |
| DGLD | Gold Bullion | -3 | -0.0010 | 0.0139 |
| IYE | Oil & Gas | 1 | 0.0000 | 0.0049 |
| DDG | Oil & Gas | -1 | -0.0008 | 0.0118 |
| DIG | Oil & Gas | 2 | -0.0005 | 0.0044 |
| DUG | Oil & Gas | -2 | -0.0018 | 0.0087 |
| DBO | WTI Crude Oil | 1 | 0.0000 | 0.0070 |
| UCO | WTI Crude Oil | 2 | -0.0006 | 0.0135 |
| SCO | WTI Crude Oil | -2 | -0.0016 | 0.0132 |
| UWTI | WTI Crude Oil | 3 | -0.0008 | 0.0147 |
| DWTI | WTI Crude Oil | -3 | -0.0017 | 0.0178 |
| IYM | Building Materials | 1 | 0.0000 | 0.0020 |
| SBM | Building Materials | -1 | -0.0004 | 0.0065 |
| UYM | Building Materials | 2 | -0.0005 | 0.0062 |
| SMN | Building Materials | -2 | -0.0022 | 0.0149 |

Table 2: Mean $\mu$ and standard deviation $\sigma$ of the 1-day tracking error by commodity.

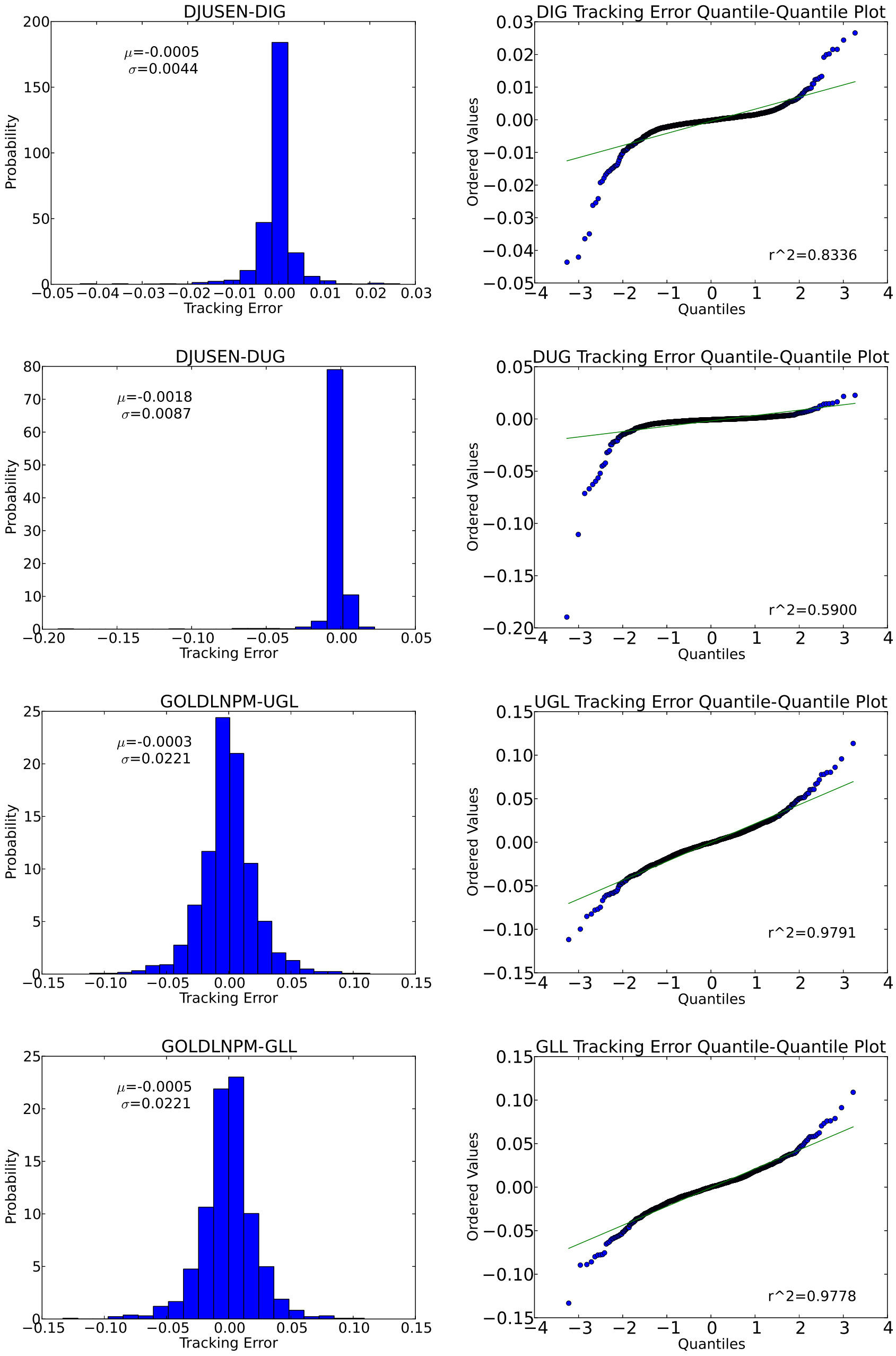

Fig. 6: Histograms and QQ plots of 1-day tracking errors for DIG, DUG ($\beta = \pm 2$, oil & gas); UGL, GLL ($\beta = \pm 2$, gold) from top to bottom.

For the 23 LETFs in Table 2, we compute the mean $\mu$ and standard deviation $\sigma$ for the tracking errors using available price data during the period Dec 2008 to May 2013. For all these funds, the mean 1-day tracking error has $\mu \approx 0$, ranging from 0% to -0.27%.

Therefore, all these LETFs on average successfully replicate the stated multiple $\beta$ of the daily reference return, with a slight negative bias. In fact, many LETFs even continued to replicate returns over periods as long as 10 days. However, as the holding time increases, the average tracking error grows more negative, so that the LETF in fact underperforms its intended goal over longer holding periods (see Figure 6).

Interestingly, the tracking errors for the silver and gold LETFs (AGQ, ZSL ($\beta = \pm 2$, silver); UGL, GLL ($\beta = \pm 2$, gold)) in Table 2 have $\sigma$ several magnitudes higher than $\mu$. For example, AGQ ($\beta = 2$, silver) has a tracking error $\sigma$ of 5% compared to a $\mu$ of 0.01%. In other words, these four LETFs, while they might track their references well on average, may also exhibit positive and negative deviations over 1-day holding periods as well. These observations are consistent with the regressions in Figures 3 and 4, where UGL and GLL ($\beta = \pm 2$, gold) show significant 1-day tracking errors. On the other hand, the non-leveraged gold and silver bullion ETFs, GLD and SLV, have almost no tracking error $\sigma \approx 0$, because they hold the underlying bullion according to their prospectuses. Since many investors use these ETFs to gain leveraged exposure to commodities, they should be aware of the large variance of the associated tracking errors.

In Figure 6, we show the histogram for the tracking error for each ETF along with a quantile-quantile plot to illustrate the distribution. For DIG and DUG ($\beta = \pm 2$, oil & gas), the quantile-quantile plot shows that the tracking error distribution is not quite normal, and has a large negative tail, so that the commodity LETF tracking error is negatively biased even for the shortest possible holding period of one day. On the other hand, for UGL, GLL ($\beta = \pm 2$, gold) the distribution appears to be normal with $R^2$ close to 98%. However, as noted in Table 2, the tracking errors for UGL and GLL ($\beta = \pm 2$, gold) also have a very large variance.

Next, we examine the horizon effect of tracking errors. Figure 7 indicates that higher leveraged ETFs tend to have more negative average tracking errors, which appear to be decreasing linearly over longer holding periods. In addition, negative leveraged LETFs have a more negative average tracking error than their positive counterparts. For example, in Figure 7, GLL ($\beta = -2$, gold) has a lower slope than UGL ($\beta = 2$, gold) even though they have the same absolute value of leverage ratio $|\beta|$. Furthermore, with few exceptions,the average tracking error is most negative when $\beta = -3$ followed by $\beta = 3, -2, 2, -1, 1$. Thus, there is a higher holding horizon punishment for buying short than long LETFs.

Our analysis of the tracking error distribution reveals several characteristics of the tracking error defined in (1). Over a very short holding period, most LETFs perform close to their objectives stated in their prospectuses. Nevertheless, the realized tracking error varies over time, and can be positive or negative. For gold and silver LETFs, the tracking error is more volatile. Moreover, the magnitude of the mean tracking error depends heavily on the $\beta$ of the LETF, with bear LETFs suffering a higher penalty than bull LETFs.

## 3 Incorporating Realized Variance into Tracking Error Measurement

As is well known in the industry (see [2, 3]), the price dynamics of an LETF depends on the realized variance of the reference index. This leads us to incorporate the realized variance in measuring the performance of an LETF. We run a regression analysis based on empirical LETF and reference prices that incorporates the realized variance as an independent variable. We then derive a realized effective fee associated with each LETF and analyze the realized price behavior relative to a theoretical benchmark to better quantify the over/under-performance.

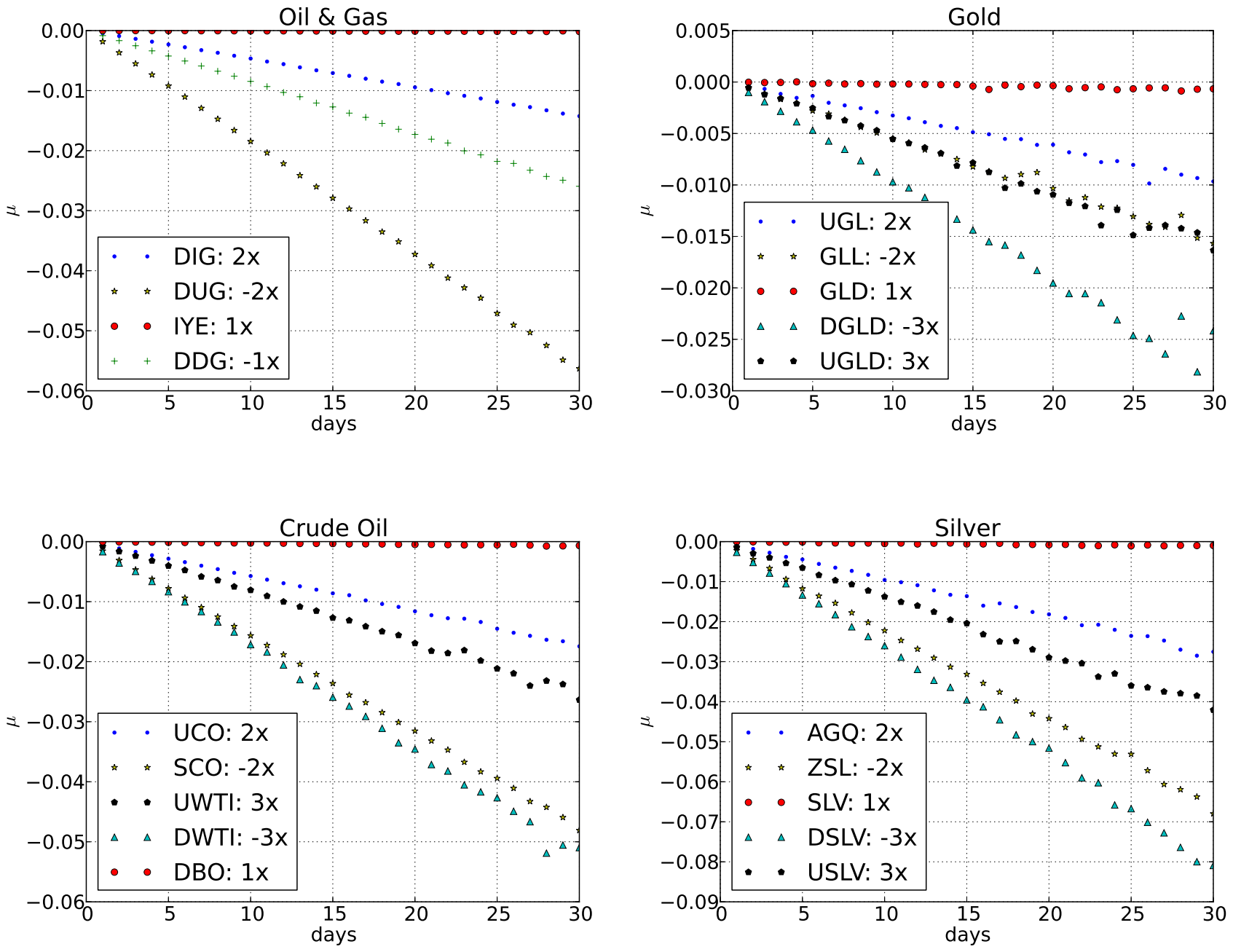

Fig. 7: A plot of no. of days vs the mean tracking error arranged by commodities tracked. From top left to bottom right: US Oil & Gas, Gold, Crude Oil,and Silver. As the holding period increases, the average tracking error becomes more negative as well.

## 3.1 Model for the LETF Price

Let $S_t$ be the price of the reference index, and $L_t$ be the price of the LETF at time $t$. Also denote $f$ as the expense rate, $r$ as the interest rate and $\beta$ as the leverage ratio. Assume the reference asset follows the SDE

$$\frac{dS_t}{S_t} = \mu_t dt + \sigma_t dW_t, \quad t \geq 0, \tag{3}$$

with stochastic drift $(\mu_t)_{t\geq 0}$ and volatility $(\sigma_t)_{t\geq 0}$. For our analysis herein, we assume a general diffusion framework, but do not need to specify a parametric model. Many well-known models, including the CEV, Heston, and exponential Ornstein-Uhlenbeck models, fit within the above framework.

A long $\beta$-LETF $L$ can be constructed through a dynamic portfolio. Specifically, the portfolio at time $t$ consists of the cash amount $\$\beta L_t$ invested in the reference index $S_t$, while $\$(\beta - 1)L_t$ is borrowed at the positive risk free rate $r$. As a result, the LETF satisfies the SDE

$$dL_t = L_t \beta \frac{dS_t}{S_t} - L_t((\beta - 1)r + f)dt. \tag{4}$$

Solving the SDE, the log-return of the LETF is given by

$$\ln \frac{L_t}{L_0} = \beta \ln \frac{S_t}{S_0} + \frac{\beta - \beta^2}{2} V_t + ((1-\beta)r - f)t, \tag{5}$$

where

$$V_t = \int_0^t \sigma_s^2 ds \tag{6}$$

is the realized variance of $S$ accumulated up to time $t$. Therefore, under this general diffusion model, the log-return of the LETF is proportional to the log-return of the reference index by a factor of $\beta$, but also proportional to the variance by a factor of $\frac{\beta-\beta^2}{2}$. The latter factor is negative if $\beta \notin (0,1)$, which is true for every LETF traded on the market. Also, the expense fee $f$ reduces the return of the LETF.

Our regression analysis will focus on testing the functional form (5). We observe from (5) that the functional form of $L_t$ in terms of $S_t$ and $V_t$ holds for any parametric model within the diffusion framework in (3). Considering the daily LETF returns, we set $\Delta t = \frac{1}{252}$ as one trading day. Let $R_t^S$ be the daily return of the reference index at time $t$. At any time $t$, the $n$-day log-returns of an LETF follows

$$\ln \frac{L_{t+n\Delta t}}{L_t} = \beta \ln \frac{S_{t+n\Delta t}}{S_t} + \frac{\beta - \beta^2}{2} V_t^{(n)} + ((1-\beta)r - f)n\Delta t, \tag{7}$$

$$V_t^{(n)} = \sum_{i=0}^{n-1} (R_{t+i\Delta t}^S - \bar{R}_t^S)^2, \quad \bar{R}_t^S = \frac{1}{n} \sum_{i=0}^{n-1} R_{t+i\Delta t}^S. \tag{8}$$

This serves as a benchmark process for our subsequent analysis.

## 3.2 Regression of Empirical Returns

The log-return equation (7) suggests a regression with two predictors: the log-returns and the realized variance of the reference over $n$-days. This results in the linear model

$$\ln \frac{L_t}{L_0} = \hat{\beta} \ln \frac{S_t}{S_0} + \hat{\theta} V_t + \hat{c} + \varepsilon, \tag{9}$$

where $\hat{c}$ is a constant intercept to be determined, and $\varepsilon \sim N(0, \sigma^2)$ is independent of $(S_t)_{t\geq 0}$.

In Table 3, we summarize the estimated $\hat{\theta}$ from our regression with holding periods of 30 days. Again, we use price data from disjoint periods to calculate returns. The realized variance is calculated using the inter-period returns (30 days). The choice of 30-day periods gives us sufficient points to compute the realized variance while providing enough disjoint periods during the period Dec 2008-May 2013 to perform a regression. A longer price history would certainly have helped in balancing this tradeoff, but all these commodity LETFs were introduced only in the past five years.

Our empirical analysis confirms several aspects of our theoretical model in (5) and provides explanations in cases where there is discrepancy. The theoretical value of $\theta$ according to (5) is given by $\frac{\beta-\beta^2}{2}$. Table 3 shows that the estimator $\hat{\theta}$ is typically in the neighborhood of $\theta$, its theoretical value. For example, SCO ($\beta = -2$, crude oil) has $\hat{\theta} = 2.93$ versus a theoretical $\theta$ of 3. In addition, the non-leveraged ETFs all have $\hat{\theta}$ close to 0, suggesting that realized variance does not play an important role in its price process, as predicted.

However, some LETFs have $\hat{\theta}$ diverging significantly from $\theta$. For example, the $\hat{\theta}$ for UGL ($\beta = 2$, gold) differs from its theoretical value by a factor of 114% even with a regression $R^2$ of 99%.

| **LETF** | **Underlying** | $\beta$ | $\hat{\theta}$ | $\theta$ | $r^2$ | $r^2_{x\mid y}$ | $r^2_{y\mid x}$ |
|---|---|---|---|---|---|---|---|
| SLV | Silver Bullion | 1 | 0.11 | 0 | 0.9799 | 0.9503 | 0.0078 |
| AGQ | Silver Bullion | 2 | -1.31 | -1 | 0.9885 | 0.9751 | 0.3892 |
| ZSL | Silver Bullion | -2 | -3.27 | -3 | 0.9995 | 0.9988 | 0.7514 |
| USLV | Silver Bullion | 3 | -2.24 | -3 | 0.9995 | 0.9988 | 0.7514 |
| DSLV | Silver Bullion | -3 | -6.94 | -6 | 0.9994 | 0.9989 | 0.9654 |
| GLD | Gold Bullion | 1 | -0.14 | 0 | 0.9898 | 0.9791 | 0.0064 |
| UGL | Gold Bullion | 2 | -2.44 | -1 | 0.9934 | 0.9867 | 0.2900 |
| GLL | Gold Bullion | -2 | -0.96 | -3 | 0.9914 | 0.9828 | 0.0417 |
| UGLD | Gold Bullion | 3 | -2.38 | -3 | 0.9982 | 0.9955 | 0.6355 |
| DGLD | Gold Bullion | -3 | -6.26 | -6 | 0.9846 | 0.9685 | 0.0809 |
| IYE | Oil & Gas | 1 | -0.06 | 0 | 0.9988 | 0.9965 | 0.1905 |
| DDG | Oil & Gas | -1 | -0.99 | -1 | 0.8866 | 0.7662 | 0.2342 |
| DIG | Oil & Gas | 2 | -1.11 | -1 | 0.9996 | 0.9989 | 0.9498 |
| DUG | Oil & Gas | -2 | -3.31 | -3 | 0.9884 | 0.9769 | 0.8873 |
| DBO | WTI Crude Oil | 1 | -0.02 | 0 | 0.9992 | 0.9981 | 0.0035 |
| UCO | WTI Crude Oil | 2 | -1.15 | -1 | 0.9987 | 0.9972 | 0.7747 |
| SCO | WTI Crude Oil | -2 | -2.93 | -3 | 0.9987 | 0.9975 | 0.9619 |
| UWTI | WTI Crude Oil | 3 | -2.14 | -3 | 0.9974 | 0.9939 | 0.6218 |
| DWTI | WTI Crude Oil | -3 | -7.25 | -6 | 0.9974 | 0.9939 | 0.6218 |
| IYM | Building Materials | 1 | 0.03 | 0 | 0.9996 | 0.9987 | 0.0495 |
| SBM | Building Materials | -1 | -0.98 | -1 | 0.9970 | 0.9920 | 0.5446 |
| UYM | Building Materials | 2 | -1.10 | -1 | 0.9997 | 0.9993 | 0.9380 |
| SMN | Building Materials | -2 | -3.59 | -3 | 0.9613 | 0.9221 | 0.5301 |

Table 3: $\hat{\theta}$ vs. $\theta$, estimated from 30-day multi-variable regression of returns, with a partial correlation table. $r^2_{y|x}$ stands for the marginal predictive power of adding the realized variance ($y$) into the model, holding constant the predictive power of the reference index returns ($x$). Similar definition for $r^2_{x|y}$. Data from Dec 2008-May 2013.

We attribute the deviation of $\hat{\theta}$ from $\theta$ in our regression to the collinearity effect of the two predictors ($\ln \frac{S_t}{S_0}$ and $V_t$). Of course $\ln \frac{S_t}{S_0}$ and $V_t$ cannot be independent observations, since $V_t$ depends on the price path process of $S_t$, the reference index. In general, the reference returns and the realized variance are negatively correlated. When the realized variance is high, it is likely the reference has suddenly dropped in value. When the realized variance is low, it usually implies a period of steady positive growth for the reference. Thus, the multi-collinearity effect is responsible for shifting predictive power among the different predictor variables. In order to measure the magnitude of the collinearity effect and the contribution of each correlated predictor variable, we compute the coefficients of partial determination for our regression model.

The factor $r^2_{y|x}$ which measures the marginal predictive power of adding the realized variance into the model. As $r^2_{y|x}$ increases, $\hat{\theta}$ becomes closer to $\theta$, suggesting a larger dependence of LETF returns on realized variance during holding periods of high volatility. For example, for the 3 LETFs DIG ($\beta = 2$, oil & gas), SCO ($\beta = -2$, crude oil), and UYM ($\beta = 2$, building materials) all have $r^2_{y|x}$ over 90%. Their estimated $\hat{\theta}$ is similarly very close to the theoretical $\theta$, never differing by more than 10%. However, for non-leveraged ETFs, the realized variance has minimal added predictive power in the model. For those ETFs, we observe $\hat{\theta} \approx 0$. For example, SLV ($\beta = 1$, silver), GLD ($\beta = 1$, gold), and DBO ($\beta = 1$, crude oil) all have $r^2_{y|x} \approx 0$, and they subsequently have $\hat{\theta} \approx 0$. In addition, $r^2_{x|y}$, which is the marginal predictive power of adding the log-returns of the reference into our regression model, is always very high, indicating that the log-returns of the reference affect the LETF prices the most, but that the realized variance is still important for predictive power, especially when leverage and the holding period is high.

## 3.3 Realized Effective Fee

In Figure 8, we show three empirical price paths: the LETF log-returns, the benchmark process defined in (5), and $\beta$ times the reference index log-returns. As we can see, the value erosion due to realized variance (volatility decay) starts to play a significant role in determining LETF prices as the holding time increases. The path associated with $\beta$ times the reference log-returns dominates the LETF log-returns after about 1 month of holding. After about 1 year, the benchmark which incorporates volatility decay more closely models the empirical LETF log-returns. For example, after 6 months of holding, SCO ($\beta = -2$, crude oil) diverges from $\beta$ times the reference, illustrating the effects of volatility decay.

However, there are also some strong deviations from the predictions given by the benchmark, which compound as the holding time increases. This causes the LETF to underperform even after the volatility decay is accounted for. For example, DUG's ($\beta = -2$, oil & gas) empirical returns begin to trail its benchmark significantly around 2009. Therefore, the volatility decay cannot explain all the LETF underperformance.

We are therefore motivated to quantify the over/under-performance of the LETFs after observing deviations from the benchmark in Figure 8. We introduce the concept of *realized effective fee* (REF) as the effective deduction rate charged by the LETF provider over the frictionless dynamic portfolio from which the LETF is constructed in Section 3.1. For a holding interval $[0,t]$, the corresponding REF is defined by

$$\hat{f}_t = (1-\beta)r - \frac{\ln\frac{L_t}{L_0} - \beta\ln\frac{S_t}{S_0} - \frac{\beta-\beta^2}{2}V_t}{t}. \tag{10}$$

Since for each LETF, $L_t$, $S_t$, $V_t$, $\beta$, and $r$ are all known, we can calculate the REF $\hat{f}_t$ for any LETF over a given holding period $[0,t]$ using historical prices. We remark that the REF, which is indexed by time $t$, depends on the selected holding horizon.

In many cases, the REF is seen to be much larger than the fund's advertised fee, indicating significant underperformance. Out of the 23 commodity LETFs, 2 have negative implied costs, so that the fund overperforms by the end of the five year period Dec 2008 to May 2013. If the REF exceeds the advertised fee, then the investor effectively pays an extra price for the opportunity to invest in the LETF. As a general trend, the bear LETFs tend to charge higher REFs than bull LETFs with the same magnitude of leverage $|\beta|$. For example, USLV ($\beta = 2$, silver) has a REF of 93 bps, while DSLV ($\beta = -2$, silver) has an REF of 504 bps over the period Dec 2008-May 2013. The two highest REFs correspond to DUG ($\beta = -2$, oil & gas) and SMN ($\beta = -2$, building materials), whose REFs are 1134 bps and 1625 bps respectively. Figure 8 illustrates that DUG ($\beta = -2$, oil & gas) drastically underperforms the benchmark, thereby realizing a high REF. Notice that in both cases, however, DUG and SMN's bull counterparts DIG ($\beta = 2$, oil & gas) and UYM ($\beta = 2$, building materials) respectively)display a negative REF, indicating overperformance during the same period. It is possible that as the reference trends upwards for a long period of time, the bear LETF will underperform, while the bull LETF will overperform.

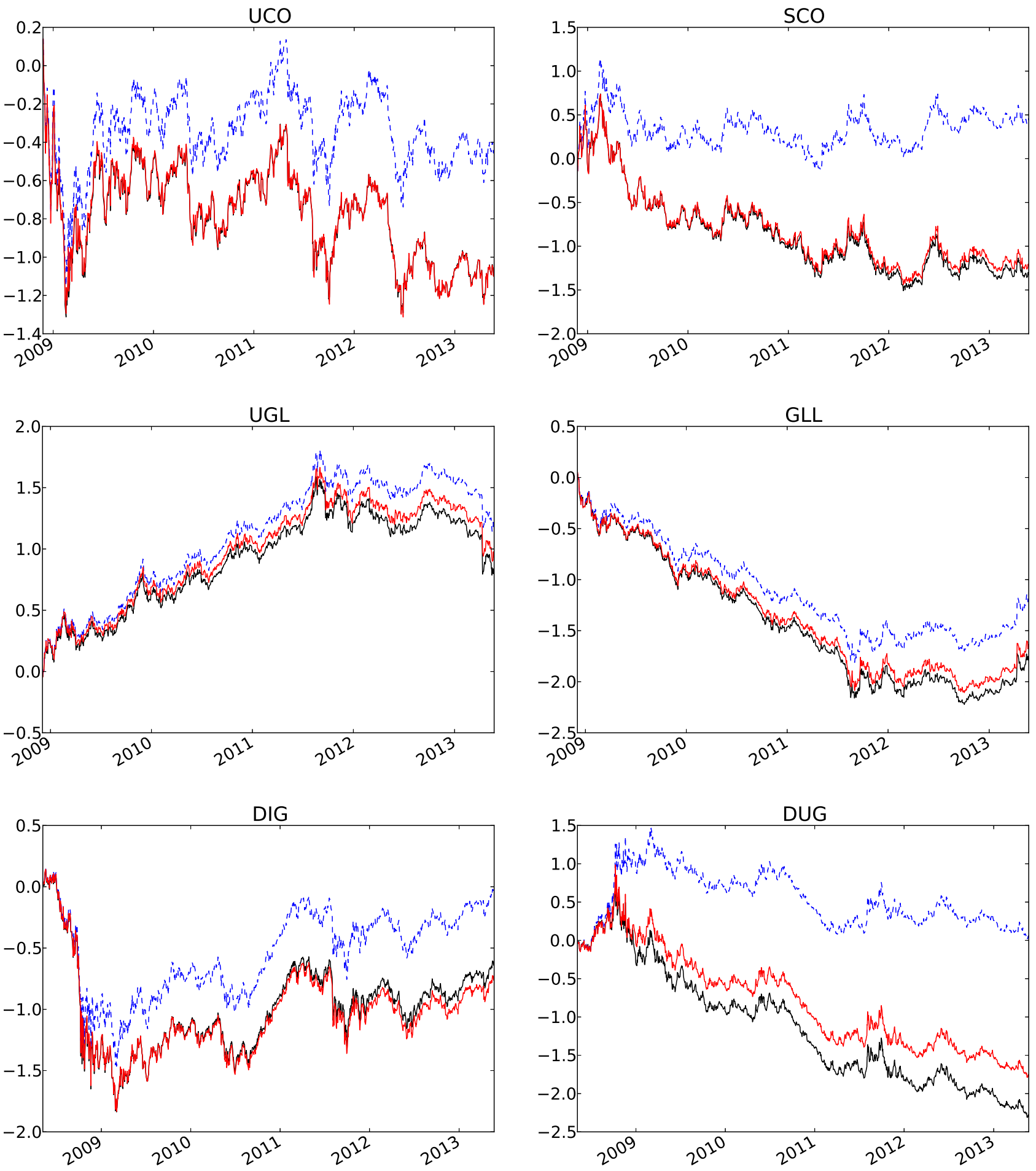

Fig. 8: Cumulative empirical log-returns of the LETF (solid dark) vs benchmark (solid light) and $\beta$ times reference (dashed light), from Dec 2008-May 2013. From top left to bottom right: UCO, SCO (crude oil); UGL, GLL (gold); DIG, DUG (building materials). UCO, UGL, and DIG have $\beta = 2$ while SCO, GLL, and DUG have $\beta = -2$.

## 4 A Static LETF Portfolio

Taking advantage of the volatility decay, a well-known trading strategy used by practitioners involves shorting a $\pm\beta$ pair of LETFs with the same reference, as discussed in [2, 7, 9, 11]. Since the LETFs have opposite daily returns on the same reference index, the portfolio has very little exposure to the reference as long as the holding period is sufficiently short. With this strategy, the volatility decay can help generate profit, which is the intuition of many practitioners. However, the portfolio is exposed to risk during periods of low volatility and high trending, as well as tracking errors. In this section, we describe an extension of this trading strategy by allowing the positive and negative leverage ratios to differ. We determine the portfolio weights to approximately eliminate the dependence on the reference. We show that the resulting portfolio is long volatility. For a number of LETF pairs, we find from empirical data that on average the strategy is profitable with enormous tail risk.

| LETF | Underlying | $\beta$ | **Prospectus Fee** (bps) | **Realized Effective Fee** (bps) |
|---|---|---|---|---|
| SLV | Silver Bullion | 1 | 50 | 96 |
| AGQ | Silver Bullion | 2 | 95 | 524 |
| ZSL | Silver Bullion | -2 | 95 | 567 |
| USLV | Silver Bullion | 3 | 165 | 93 |
| DSLV | Silver Bullion | -3 | 165 | 504 |
| GLD | Gold Bullion | 1 | 40 | 48 |
| UGL | Gold Bullion | 2 | 95 | 343 |
| GLL | Gold Bullion | -2 | 95 | 406 |
| UGLD | Gold Bullion | 3 | 135 | 139 |
| DGLD | Gold Bullion | -3 | 135 | 521 |
| IYE | Oil & Gas | 1 | 48 | 50 |
| DDG | Oil & Gas | -1 | 95 | 953 |
| DIG | Oil & Gas | 2 | 95 | -142 |
| DUG | Oil & Gas | 2 | 95 | 1134 |
| DBO | WTI Crude Oil | 1 | 75 | 56 |
| UCO | WTI Crude Oil | 2 | 95 | 84 |
| SCO | WTI Crude Oil | -2 | 95 | 321 |
| UWTI | WTI Crude Oil | 3 | 135 | 3 |
| DWTI | WTI Crude Oil | -3 | 135 | 549 |
| IYM | Building Materials | 1 | 48 | 11 |
| SBM | Building Materials | -1 | 95 | 456 |
| UYM | Building Materials | 2 | 95 | -204 |
| SMN | Building Materials | -2 | 95 | 1625 |

Table 4: Comparison of the official fee for the LETF charged on the fund prospectus and the REF calculated using 5 years of price data (December 2008-May 2013) for the LETF and reference (see (10)). We set $r = 69.1$ bps, the annualized LIBOR rate.

We now construct a weighted portfolio which is short the LETF with leverage ratio $\beta_+ > 0$ and short another LETF with leverage ratio $\beta_- < 0$. We emphasize that both LETFs having the same reference, but that $\beta_+$ and $|\beta_-|$ may differ. We hold fraction $\omega \in (0,1)$ of the portfolio in the $\beta_+$-LETF and $(1-\omega)$ of the portfolio in the $\beta_-$-LETF. At time $T$, the normalized return from this strategy is

$$\mathscr{R}_T = 1 - \omega \frac{L_T^+}{L_0^+} - (1-\omega)\frac{L_T^-}{L_0^-}. \tag{11}$$

Applying (5), $\mathscr{R}_T$ admits the expression

$$\mathscr{R}_T = 1 - \omega \left(\frac{S_T}{S_0}\right)^{\beta_+} \exp(\Gamma_T^+) - (1-\omega)\left(\frac{S_T}{S_0}\right)^{\beta_-} \exp(\Gamma_T^-), \tag{12}$$

where

$$\Gamma_T^{\pm} = \frac{\beta_\pm - \beta_\pm^2}{2} V_T + ((1-\beta_\pm)r - f_\pm)T, \tag{13}$$

Here, $\beta_\pm$ and $f_\pm$ are the respective leverage ratios and fees of the two LETFs in the portfolio defined in (11). Over a short holding period such that $\frac{L_T}{L_0} \approx 1$ , one can pick an appropriate weight $\omega^*$ to approximately remove the dependence of $\mathscr{R}_T$ on $S_T$.

**Proposition 1.** *Select the portfolio weight* $\omega^* = \frac{-\beta_-}{\beta_+ - \beta_-}$. *For* $\frac{L_T}{L_0} \approx 1$, *the return from this strategy is given by*

$$\mathscr{R}_T = \frac{-\beta_-\beta_+}{2} V_T - \frac{\beta_-}{\beta_+ - \beta_-}(f_+ - f_-)T + (f_- - r)T. \tag{14}$$

*Proof.* For $\frac{L_T}{L_0} \approx 1$, we can substitute for $\frac{L_T}{L_0}$ with $\ln\frac{L_T}{L_0} + 1$ in (11). Then, we set $\omega = \omega^*$ and apply (5) to conclude (14).

The return (14) corresponding to portfolio weight $\omega^*$ reflects a linear dependence on the realized variance. In particular, the coefficient $\frac{-\beta_-\beta_+}{2}$ is strictly positive, so the strategy is effectively long volatility ($V_T$). Also, as it does not depend on $S_T$, the $\omega^*$ portfolio is $\Delta$-*neutral* as long as the reference does not move significantly. In Table 5, we summarize the coefficient of $V_T$ and the weighted portfolio $(\omega^*, 1-\omega^*)$ for different combinations of leverage ratios. Note that as long as $\beta_+ = -\beta_-$, we end up with the portfolio weight $\omega^* = \frac{1}{2}$. Also, the coefficient $\frac{-\beta_-\beta_+}{2}$ exceeds or equals to 1 except for the pair $(\beta_+, \beta_-) = (1, -1)$, and it is largest for the pair $(\beta_+, \beta_-) = (3, -3)$.

| $(\beta_+, \beta_-)$ | $\omega^*$ | $\frac{-\beta_-\beta_+}{2}$ |
|---|---|---|
| $(1,-1)$ | 1/2 | 1/2 |
| $(1,-2)$ | 2/3 | 1 |
| $(1,-3)$ | 3/4 | 3/2 |
| $(2,-1)$ | 1/3 | 1 |
| $(2,-2)$ | 1/2 | 2 |
| $(2,-3)$ | 3/5 | 3 |
| $(3,-1)$ | 1/4 | 3/2 |
| $(3,-2)$ | 2/5 | 3 |
| $(3,-3)$ | 1/2 | 9/2 |

Table 5: Table of $(\beta_+, \beta_-)$ pairs vs $\omega^*$ the weight of the $\beta_+$ portfolio, and $\frac{-\beta_-\beta_+}{2}$ the dependence of the strategy on $V_t$ (see Prop. 1).

We now backtest the $\omega^*$ strategy from Prop. 1 as follows. For each LETF pair, we short \$0.5 of the $\beta_+$-LETF and \$0.5 of the $\beta_-$-LETF with $\beta_+ = -\beta_- = 2$ and hold the position for some time $T$. The normalized return $\mathscr{R}_T$ depends on the relative weights on the long/short-LETFs but not the absolute cash amounts. More generally, one can also test the strategy with different $\beta_\pm$ and $\omega^*$.

Dividing the price data from Dec 2008-May 2013 into $n$-day rolling (overlapping) periods, we calculate the returns from the strategy over each period. For every $n$-day return, we compare against the realized variance over the same period. This is illustrated in Figure 9. As a theoretical benchmark, we also plot $\mathscr{R}_T$ in (14) as a linear function. Each point (dot) on the plots represents a 5-day return, but over rolling periods the returns are not independent. In other words, the lines in Figure 9 are not generated by regression but taken from (14). We choose (14) as a benchmark because it is expected to hold *pathwise* as long as $\frac{L_T}{L_0} \approx 1$ with negligible tracking error.

We can observe from Figure 9 that the returns exhibit positive dependence on the realized variance ($V_T$). In particular, for the energy pairs (DIG-DUG ($\beta = \pm 2$, oil & gas) and UCO-SCO ($\beta = \pm 2$, crude oil)), the returns tend to be very positive when the realized variance is high. This is because the strategy captures the volatility decay as profit. Nevertheless, there is also a visible amount of noise in the returns deviating from the linear dependence on $V_T$, especially for the gold and silver pairs (UGL-GLL ($\beta = \pm 2$, gold) and AGQ-ZSL ($\beta = \pm 2$, silver), respectively). This can be partly attributed to tracking errors from both LETFs in the portfolio. Also, the $\omega^*$-strategy loses its $\Delta$-neutrality if the reference moves significantly.

While this portfolio is expected to be $\Delta$-neutral (with respect to the reference index) for small reference movements, in reality the strategy is also short-$\Gamma$. One way to see this is through Figure 10 that plots the returns against the reference index returns. Common to all four LETF pairs, when the reference return is either very positive or negative, the return of the $\omega^*$-strategy tends to be negative. As a theoretical benchmark, we also plot the normalized return equation (12) which applies even for large reference movements.

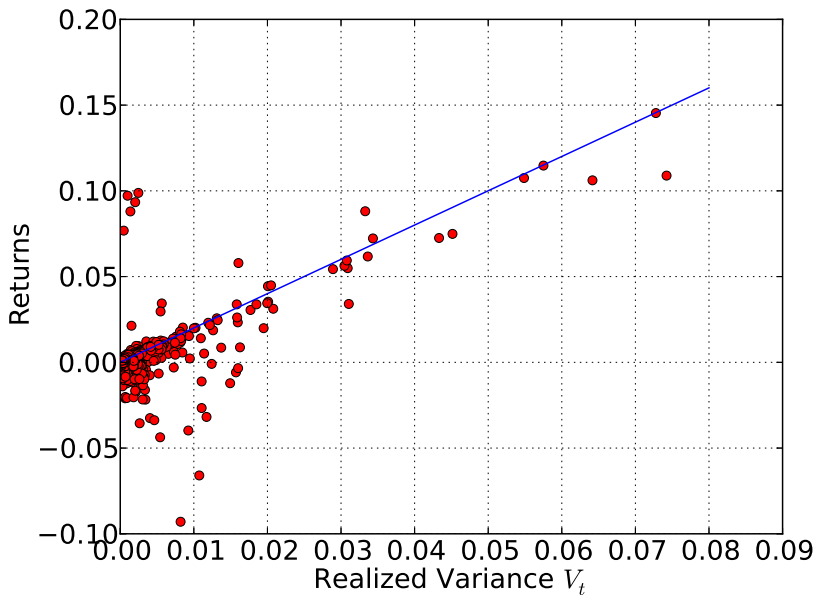

(a) DJUSEN-DIG-DUG

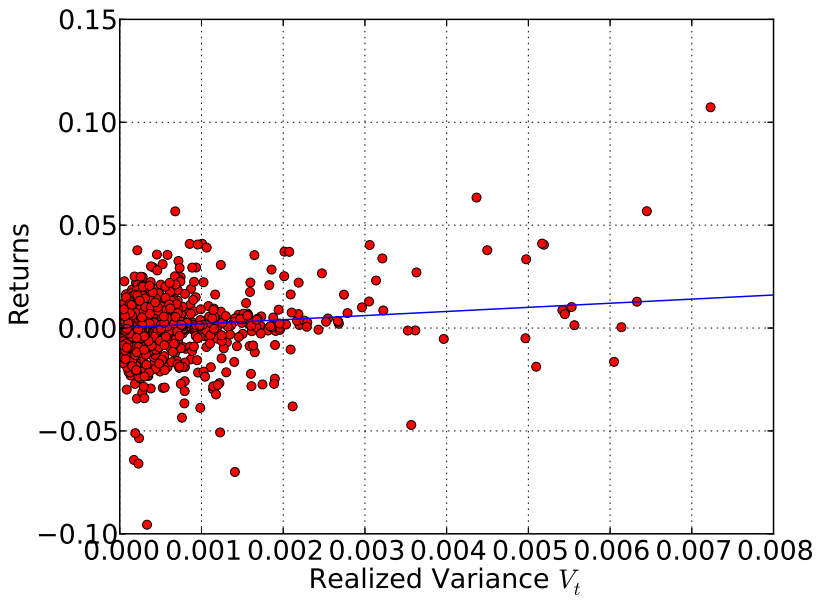

(b) GOLDLNPM-UGL-GLL

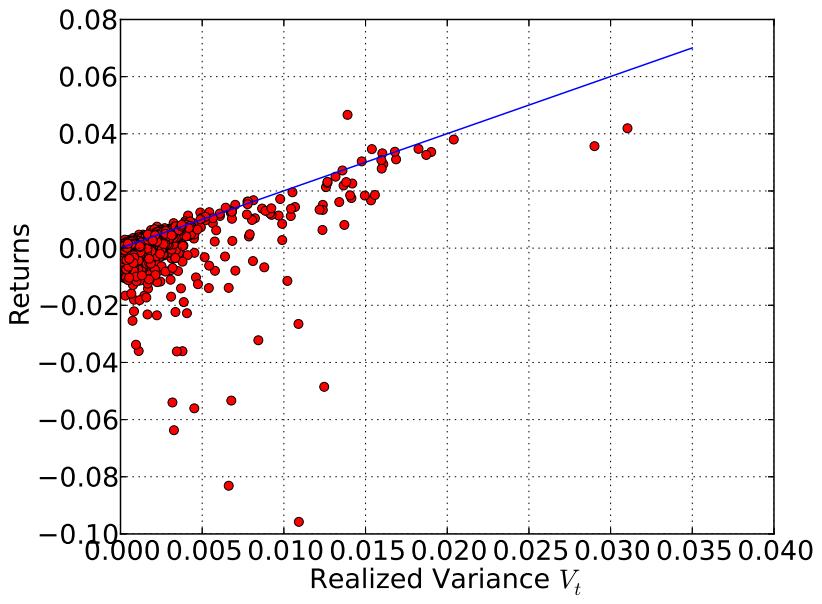

(c) DJUBSCL-UCO-SCO

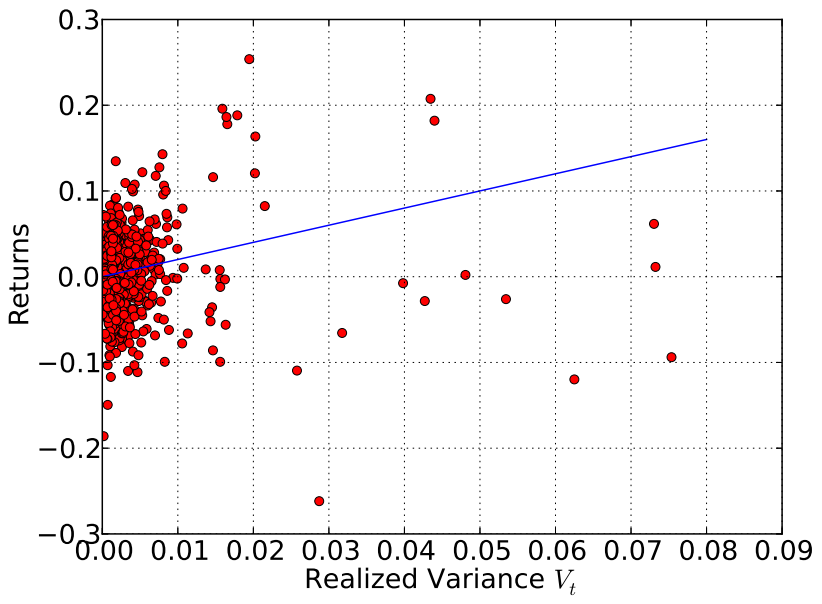

(d) SLVRLN-AGQ-ZSL

Fig. 9: Plot of trading returns vs realized variance for a double short strategy over 5-day rolling holding periods, with $\beta_{\pm} = \pm 2$ for each LETF pair. We compare with the empirical returns (circle) from the $\omega^*$ strategy with the predicted return (solid line) in Prop. 1. Trading pairs are DIG-DUG (oil & gas), UGL-GLL (gold), UCO-SCO (crude oil), AGQ-ZSL (silver).

In contrast to the energy pairs, the gold and silver pairs yield very noisy returns. This is consistent with our earlier observations from our regressions in Figures 3 and 4. For instance, both UGL and GLL ($\beta = \pm 2$, gold) show substantial tracking errors over short periods such as 5 days, and their regressed leverage ratios differ from the stated ones. On the other hand, the DIG and DUG ($\beta = \pm 2$, oil & gas) regressions in Figures 1 and 2 reflect much less tracking errors.

Furthermore, Figure 11 shows that as the holding time increases, the returns from the $\omega^*$ strategy increases as well. The performance is best for the energy pairs UCO-SCO ($\beta = \pm 2$, crude oil) and DIG-DUG ($\beta = \pm 2$, oil & gas), but more subdued for the bullion pairs UGL-GLL ($\beta = \pm 2$, gold) and AGQ-ZSL ($\beta = \pm 2$, silver). However, over longer holding periods, the $\omega^*$ portfolio may lose its $\Delta$-neutral status, thereby generating more risk as well. Although average returns from the $\omega^*$ strategy are positive, one is subject to enormous tail risk, which increases with the holding time of the static portfolio. In order to ensure that we do not subject ourselves to excessive tail risk, we should not only be sure of a high volatility environment, but we must also adjust the holding time to account for the extra risk associated with time horizon of returns.

Figure 12 gives another perspective of the $\omega^*$ strategy's dependence on realized variance. It shows the time series of the 30-day rolling returns along with the realized variance of the reference index from Dec 2008 to May 2013. We see that when the realized variance increases sharply, the strategy returns also spike sharply. For example, when DJUSEN index realized variance spikes, the DIG-DUG ($\beta = \pm 2$, oil & gas) trading pair accumulates a 30% return over a single 30-day holding period. However, when realized variance is subdued over a period of time, the $\omega^*$ returns may turn quite negative as well.

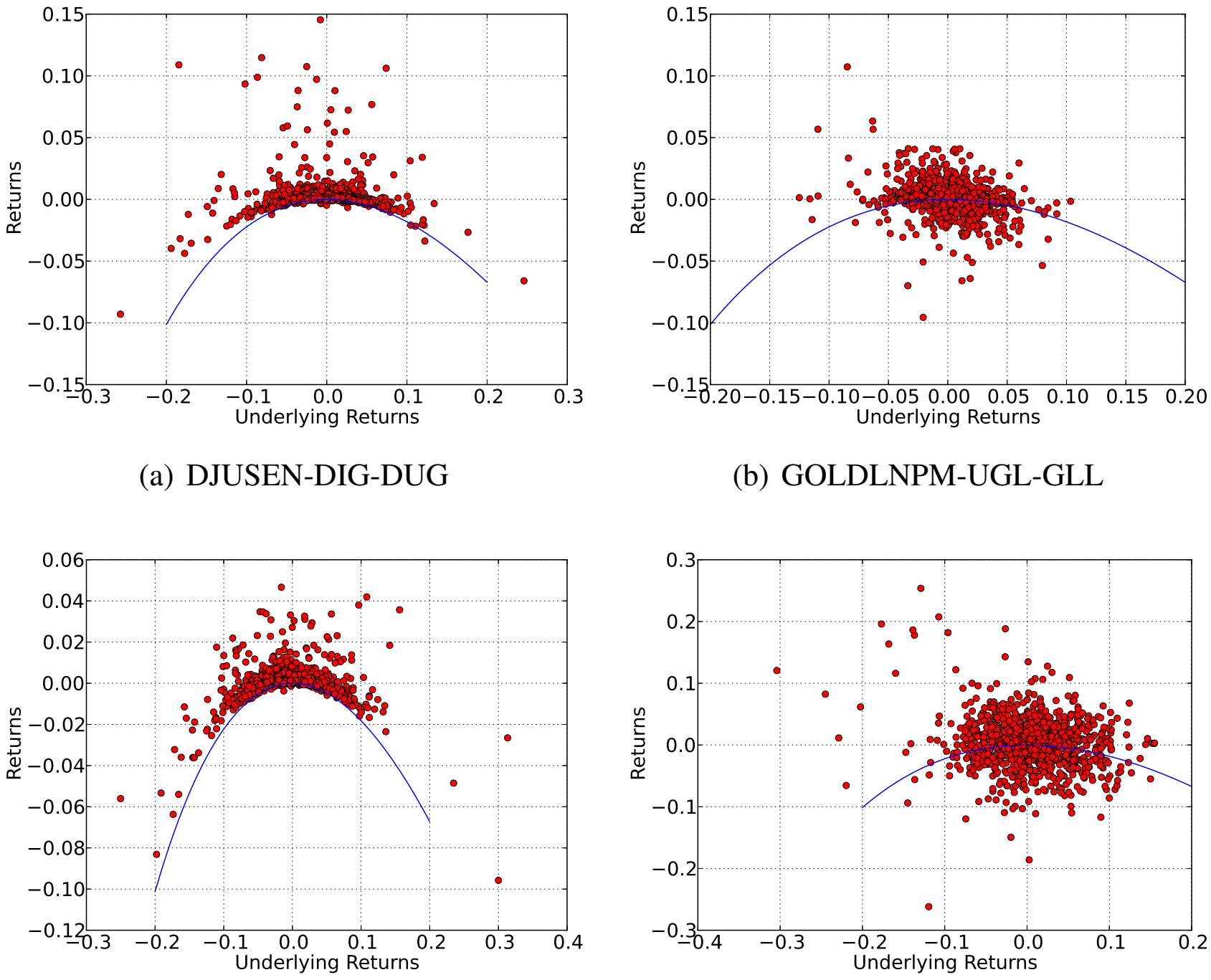

(a) DJUSEN-DIG-DUG

(b) GOLDLNPM-UGL-GLL

(c) DJUBSCL-UCO-SCO

(d) SLVRLN-AGQ-ZSL

Fig. 10: Plot of returns of reference index vs trading returns for a double short strategy over 5-day rolling, holding periods. $\beta_{\pm} = \pm 2$ for each LETF pair. We compare the empirical returns from our trading strategy (dark solid circle) with the predicted dependence on reference returns according to (12), using $\Gamma_T^{\pm} = 0$ (light solid line). Trading pairs are DIG-DUG (oil & gas), UGL-GLL (gold), UCO-SCO (crude oil), AGQ-ZSL (silver).

In summary, the double-short trading strategy studied herein is profitable on average, but it is commodity specific and subject to enormous tail risk, as seen from empirical prices. The strategy's profitability depends strongly on a high volatility from the reference index. Although longer holding times tend to enhance the average return, they also enormously increase the horizon risk. According to these findings, this strategy appears to be appealing only during times of high volatility in the reference index.

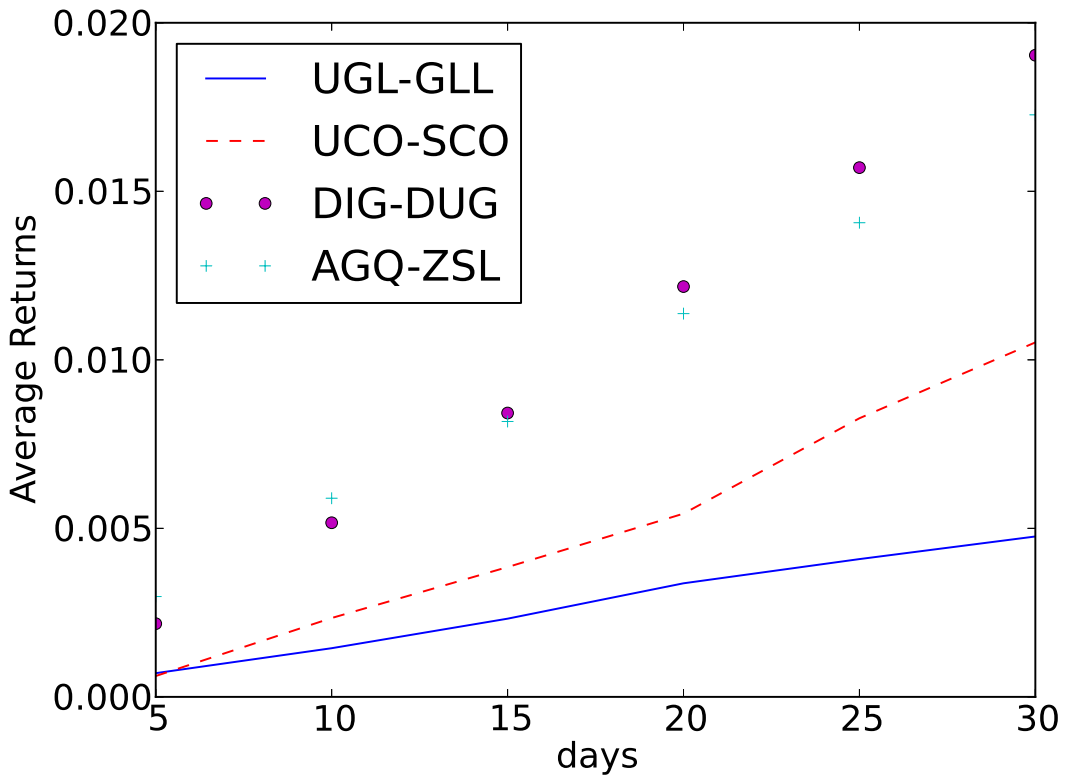

Fig. 11: Average returns from a double short trading strategy by commodity pair over no. of days holding period. $\beta_{\pm} = \pm 2$ for each LETF pair. Trading pairs are DIG-DUG (oil & gas), UGL-GLL (gold), UCO-SCO (crude oil), AGQ-ZSL (silver).

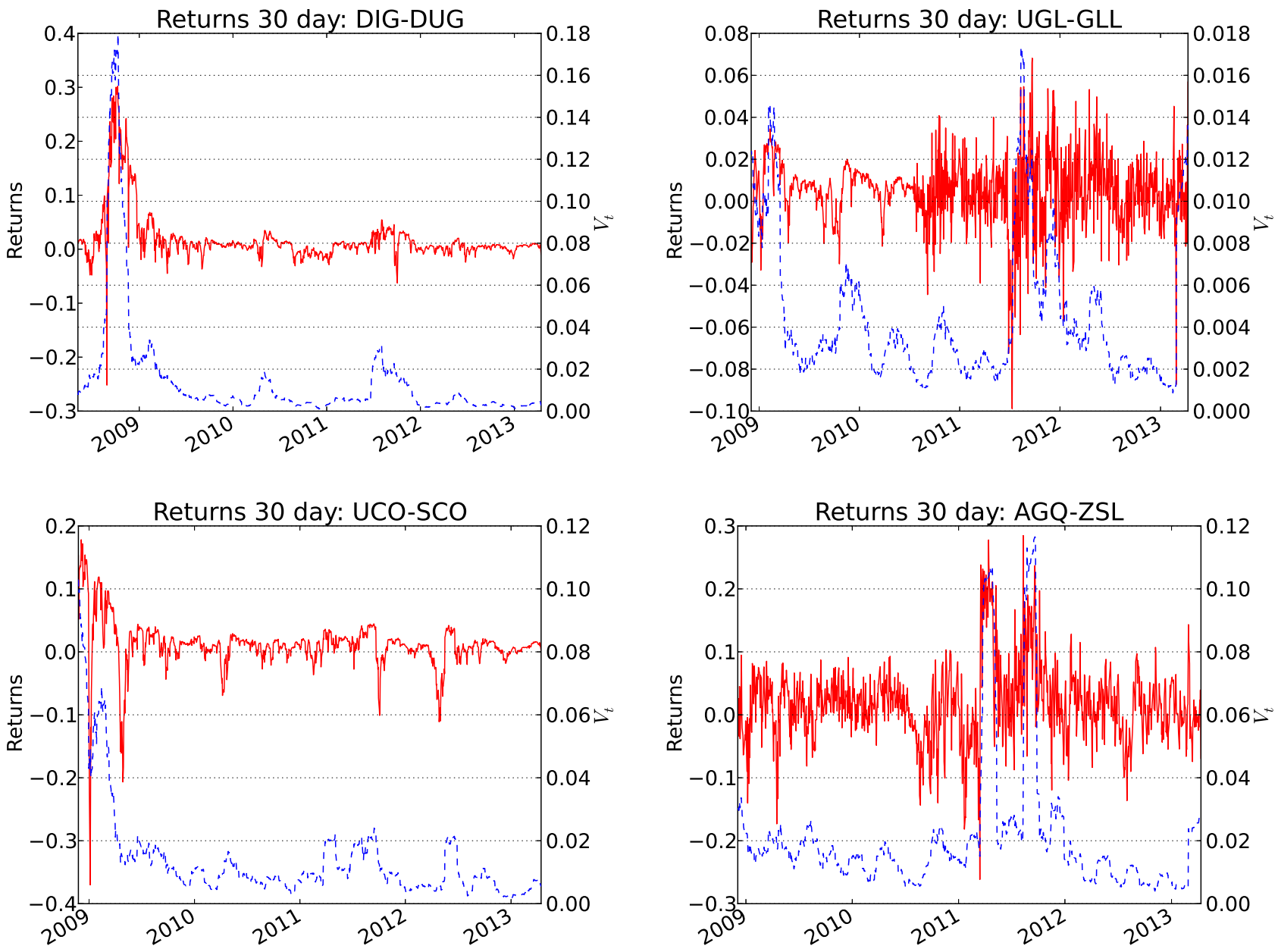

Fig. 12: Time series of returns for a double short strategy over 30-day rolling, holding periods, with $\beta_{\pm} = \pm 2$ for each LETF pair. Notice how during the periods of greatest volatility the double short strategy has the greatest return. Trading pairs are DIG-DUG (oil & gas), UGL-GLL (gold), UCO-SCO (crude oil), AGQ-ZSL (silver).

## 5 Concluding Remarks

The ETF market has continued to grow in quantity and diversity, especially in the past five years. For both investors and regulators, it is very important to understand and quantify the risks involved with various ETFs. In this paper, we have focused on commodity ETFs and their leveraged counterparts. We find that the LETF returns tend to deviate significantly from the corresponding multiple of the reference returns as the holding horizon lengthens. To study the performance of an LETF, we have applied a new benchmark process that accounts for the realized variance of the underlying. We find that many commodity LETFs still diverge, typically negatively, from this benchmark over time. These empirical observations motivate us to illustrate the over/under-performance of an LETF via the concept of realized expense fee. Based on the funds and the time periods we have studied, most commodity LETFs effectively charge significantly higher expense fees than stated on their prospectuses.

In view of LETFs' common pattern of value erosion over time, one well-known trading strategy in the industry involves statically shorting both long and short LETFs in order to capture the volatility decay as profit. We systematically study an extension of this strategy that is applicable to LETF pairs with different asymmetric leverage ratios. We analytically derive the specific weights in the LETFs so that the resulting portfolio is approximately $\Delta$-neutral, but short-$\Gamma$ as well. This strategy can potentially be quite profitable but its return can be negatively impacted by tracking errors generated by the LETFs and large movements of the reference index. These two factors both depend on the holding horizon. This should motivate future research on the horizon risk for LETF strategies. To this end, Leung and Santoli [7] study the admissible holding horizon and leverage ratio given a risk constraint. The recent papers [6, 13, 14] examine the dynamics of price spreads between ETF pairs, for example, gold vs. silver.

Our analysis herein does not assume a parametric stochastic volatility model for the underlying. It is of practical interest to investigate the price behavior of LETF under a number of well-known stochastic volatility models, such as the Heston and SABR models. On top of LETFs, there are also options written on these funds. This gives rise to the question of consistent pricing of LETF options across leverage ratios (see [1, 8]). Finally, models that capture the connection between LETFs and the broader financial market would be very useful for not only traders and investors, but also regulators.

**Acknowledgements** The authors would like to thank Scott Weiner of VelocityShares, and the participants of the 2013 Focus Program on Commodities, Energy and Environmental Finance held at Field's Institute and the 2014 Joint Mathematics Meetings in Baltimore for their helpful suggestions and comments.